\documentclass[preprint]{revtex4-1}
\usepackage{graphicx}
\usepackage{color}
\usepackage{bm}
\usepackage{braket}
\usepackage{amsmath}

\begin{document}


\title{Large Tunable Anomalous Hall Effect in the Kagom$\acute{e}$ Antiferromagnet U$_3$Ru$_4$Al$_{12}$}
\author{T. Asaba$^{1*}$, Ying Su$^{2}$, M. Janoschek$^{1,3}$, J. D. Thompson$^{1}$, S. M. Thomas$^{1}$,  E. D. Bauer$^1$, Shi-Zeng Lin$^{2}$, F. Ronning$^{4*}$}

\affiliation{
$^1$MPA-Q, Los Alamos National Laboratory, NM, 87545 USA \\
$^2$Theoretical Division, T-4 and CNLS, Los Alamos National Laboratory, NM, 87545, USA \\
$^3$Laboratory for Neutron and Muon Instrumentation, Paul Scherrer Institute, CH-5232 Villigen, Switzerland\\
$^4$Institute for Materials Science, Los Alamos National Laboratory, New Mexico, 87545, USA}
\date{\today}
\begin{abstract}
The Berry curvature in magnetic systems is attracting interest due to the potential tunability of topological features via the magnetic structure. $f$-electrons, with their large spin-orbit coupling, abundance of non-collinear magnetic structures and high electronic tunability, are attractive candidates to search for tunable topological properties. In this study, we measure anomalous Hall effect (AHE) in the distorted kagom$\acute{e}$ heavy fermion antiferromagnet U$_3$Ru$_4$Al$_{12}$. A large intrinsic AHE in high fields reveals the presence of a large Berry curvature. Moreover, the fields required to obtain the large Berry curvature are significantly different between $B \parallel a$ and $B \parallel a^*$, providing a mechanism to control the topological response in this system. Theoretical calculations illustrate that this sensitivity may be due to the heavy fermion character of the electronic structure.   These results shed light on the  Berry curvature of a strongly correlated band structure in magnetically frustrated heavy fermion materials, but also emphasize 5$f$-electrons as an ideal playground for studying field-tuned topological states.

\end{abstract}

\maketitle

{\bf Introduction} 

 The Berry curvature of a material is a property of the electronic structure that provides an anomalous transverse velocity to electrons traveling in a solid. For insulators, the integral of the Berry curvature becomes quantized leading to the notion that an electronic structure has a topology defined by its Berry curvature. This topology can lead to dramatic observables such as quantized conductance and novel boundary states \cite{klitzing1980new,konig2007quantum,hsieh2008topological,hasan2010colloquium,xu2015observation}. By tuning the topology of a system one may hope to control these properties. This has been demonstrated in several non-collinear antiferromagnets (AFM) and ferromagnets (FM), where the opposite Berry curvature from two different domains can be accessed by flipping a small applied magnetic field \cite{surgers2014large,ye2018massive,liu2018giant,nayak2016large,nakatsuji2015large}. The reversal of the Berry curvature is witnessed by the change in sign of a large intrinsic component to the anomalous Hall effect \cite{nagaosa2010anomalous}. A finite Berry curvature is a consequence of spin-orbit coupling (SOC).  SOC enables a magnetic field and/or magnetic structure to modify the electronic structure\cite{yin2018giant,mlynczak2016fermi,suzuki2016large}. This demonstrates an additional mechanism to tune the topological response of a system with a rotating magnetic field, but for typical electronic energy scales, one would expect a field of a few Tesla to be a weak perturbation.

Strong electronic correlations can further broaden the landscape of topological materials, for instance, by creating novel fractionalized particles \cite{tsui1982two,laughlin1983anomalous,witczak2014correlated,wen2017colloquium}. Importantly, it can amplify the tunability of materials through increased susceptibility to external perturbations. $f$-electron heavy fermion systems are ideal for these types of studies, as the renormalized electronic energy scales are 100 - 1000 times smaller than ordinary metals. Furthermore, strong SOC can lead to topologically non-trivial properties as well as non-collinear spin structures \cite{dzero2010topological,lai2018weyl,shirer2018dirac}. 

Here we report Hall effect measurements on the heavy fermion, non-collinear antiferromagnet U$_3$Ru$_4$Al$_{12}$. With an applied magnetic field we find a significant non-linear anomalous Hall response, which can be tuned by small rotations of the magnetic field. Theoretical calculations  reveal that this can be understood as a consequence of the magnetic field exceeding the linear response regime due to a small electronic energy scale of heavy fermion quasiparticles. Hence, the electronic structure, and consequently the Berry curvature, is significantly modified by the field strength and orientation. This work demonstrates that 5$f$-based materials are interesting model systems to investigate the tunability of the Berry curvature in the presence of strong electronic correlations.

{\bf Experiments}

\begin{figure}[h]
	\includegraphics[width=\linewidth]{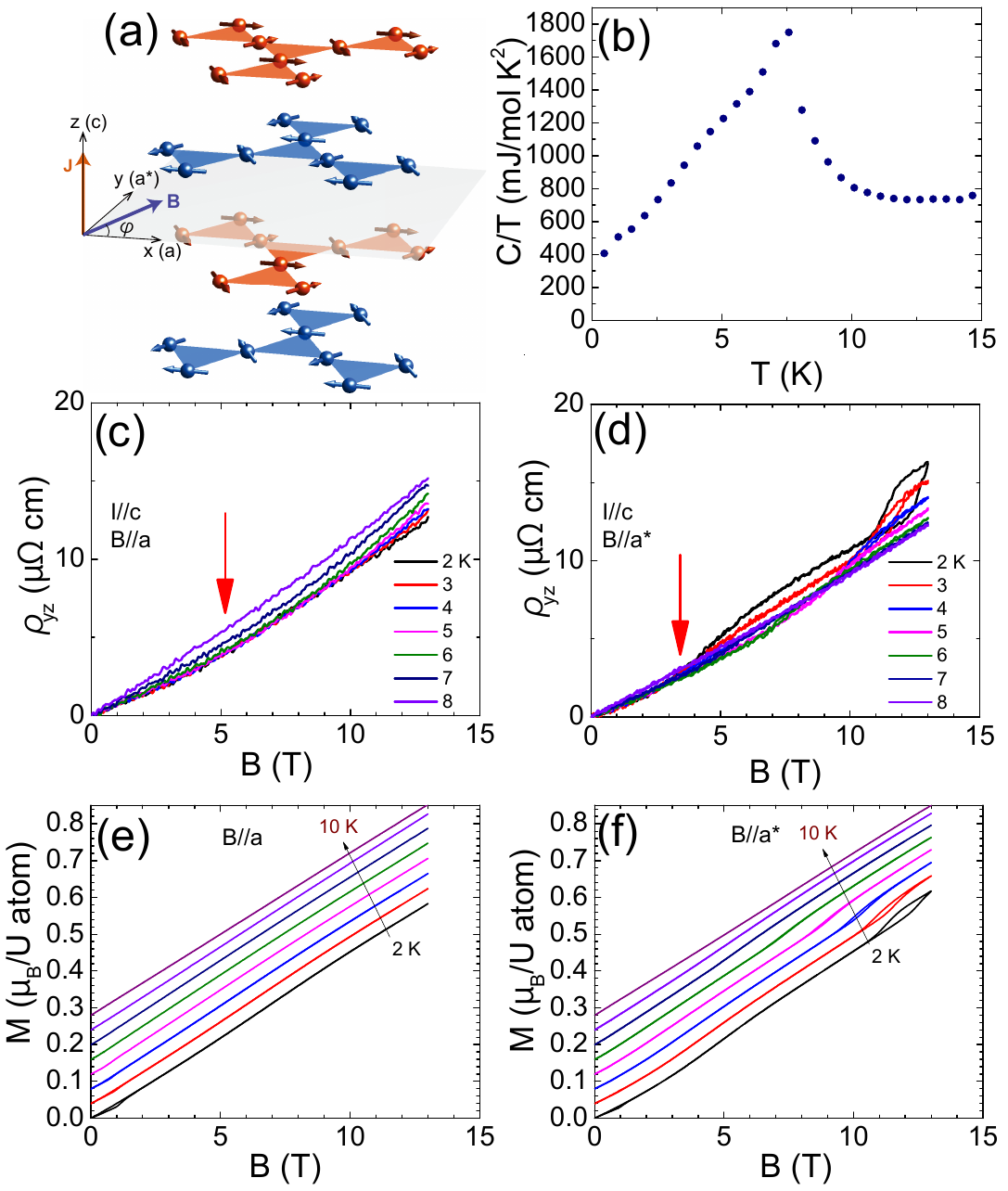}
	\caption{\label{figRyz} 
		{Magnetization and anomalous Hall resistivity at different temperatures (color online).} 
		(a) The magnetic structure of U$_3$Ru$_4$Al$_{12}$. Only uranium atoms are shown. The arrows indicate the magnetic structure determined by neutron scattering measurements \cite{troc2012single}. (b) Specific heat data of U$_3$Ru$_4$Al$_{12}$. (c)(d) Anomalous Hall resistivity with $J \parallel $c and (c) $B \parallel a$ and (d) $B \parallel a^*$. (e)(f)In-plane magnetic field dependence of magnetization with (e) $B \parallel a$ and (f) $B \parallel a^*$. The curves are offset by 0.04 $\mu_B$/U atom for clarity.
	}
\end{figure}

The heavy-fermion antiferromagnet U$_3$Ru$_4$Al$_{12}$ has a Gd$_3$Ru$_4$Al$_{12}$ type hexagonal crystal structure. Distorted kagom$\acute{e}$ nets of uranium atoms govern the magnetism. Due to the 5$f$ electrons of uranium and the frustrated kagom$\acute{e}$ geometry, the system orders at 8 K in a unique non-collinear magnetic structure shown in Fig. \ref{figRyz} (a) \cite{pasturel2009crystal}. Neutron scattering measurements show that  the spins are rotated $\pm$ 60 degrees in each triangle, resulting in a net ferromagnetic component in-plane \cite{troc2012single}. This ferromagnetic component is then canceled out by adjacent layers that have the opposite spin arrangement, forming the antiferromagnetic structure. The system possesses a large Sommerfeld  coefficient $\gamma$ of 110 mJ/mol-U K$^2$ suggesting a large effective mass within the magnetically ordered state (See Fig. \ref{figRyz}(b)). 

Our main observation is a surprising non-linearity and angular dependence in the transverse resistivity (anomalous Hall effect, AHE) in comparison to the magnetization as shown in Fig. \ref{figRyz}(c) and Fig. \ref{figRyz}(d) for $B \parallel a$ and $B \parallel a^*$, respectively.  At low fields, $\rho_{yz}$ and $M$ increase linearly with applied fields. However, an unexpected field-induced enhancement of $\rho_{yz}$ is observed in both orientations at low temperatures and low fields ($<$ 10 T), as indicated by arrows. 

Figures \ref{figRyz} (e) ($B \parallel a$) and (f) ($B \parallel a^*$) show the magnetic field dependence of magnetization at different temperatures and different field orientations. In both orientations, at low fields hysteresis loops are observed, illustrating antiferromagnetic domain reorientation (see SI). Also, the susceptibility is almost identical in $a$ and $a^*$ at low fields, indicating an isotropic response. With increasing field, both orientations show a nearly linear field dependence. An additional jump with a hysteresis loop is observed in the $B \parallel a^*$ orientation at the critical field $B_M$ =12 T at $T$ = 2 K, which is also seen in magnetoresistance (MR) data shown in the supplementary information and in magnetization and ultrasound data\cite{SI,gorbunov2019magnetic}. Thus, it is confirmed that a field-induced phase transition originates from a metamagnetic transition at $B_M$. A similar in-plane metamagnetic phase-transition anisotropy was observed in Dy$_3$Ru$_4$Al$_{12}$ \cite{gorbunov2014electronic} and Ho$_3$Ru$_4$Al$_{12}$ \cite{gorbunov2018crystal}, indicating that this property is related to the crystal electric field anisotropy and the crystal structure.

In contrast to the high-field metamagnetic transition, the field-induced non-linear anomalous Hall conductivity (AHC) behavior is not due to a phase transition. Indeed, no anomaly is observed in magnetization or specific heat data at the field where $\rho_{yz}$ is suddenly enhanced, as shown in Figs. \ref{figRyz} (c) and (d). Below, we discuss that the AHC possesses both extrinsic and intrinsic anomalous Hall contributions, and by subtracting a contribution proportional to the magnetization, we find a significant field and temperature dependent intrinsic AHE contribution. 


In general, in magnetic materials, the transverse resistivity $\rho_{yz}$ is expressed as
\begin{equation} \label{AHEeq}
\rho_{yz} = R_H B + \mu_0 R_s^{ext} M +R_{AHE}^{int}
\end{equation}
where $R_H$, $R_s^{ext}$ and $R_{AHE}^{int}$ are ordinary, extrinsic anomalous, and intrinsic anomalous Hall contributions, respectively.  The ordinary Hall contribution $R_H B$ can be determined at high temperatures and is found to be negligible \cite{SI}. Thus, the expression of $\rho_{yz}$ reduces to 
\begin{equation} \label{AHEeq2}
\rho_{yz} = \mu_0 R_s^{ext} M +R_{AHE}^{int}.
\end{equation}
The extrinsic AHE $\mu_0 R_s^{ext} M$ originates from a scattering mechanism. The intrinsic AHE $R_{AHE}^{int}$ originates from the Berry curvature. Because the magnetic structure shown in Fig. \ref{figRyz} (a) is symmetric with the product of the time reversal operation and the inversion symmetry operation $\mathcal{P} \times \mathcal{T}$, $R_{AHE}^{int}$  vanishes at zero field. In low fields when $\rho_{yz}$ and $M$ are proportional we cannot distinguish between the intrinsic and extrinsic responses. However, we note that Onoda $\it{et}$ $\it{al.}$ suggest that the  AHC is dominated by the intrinsic component when $\rho$ $>$ 10-100 $\mu\Omega$cm for rare earth compounds  \cite{onoda2008quantum}. Since the resistivity of U$_3$Ru$_4$Al$_{12}$ is larger than 300 $\mu\Omega$cm, the anomalous Hall effect could be solely attributed to the intrinsic contribution. If this is the case, the intrinsic AHC would reach about 100 (280) $\Omega^{-1} $cm$^{-1}$  with B$\parallel$ a* (B$\parallel$ c) (for B$\parallel$c data, see SI \cite{SI}).

\begin{figure}[h]
	\includegraphics[width=\linewidth]{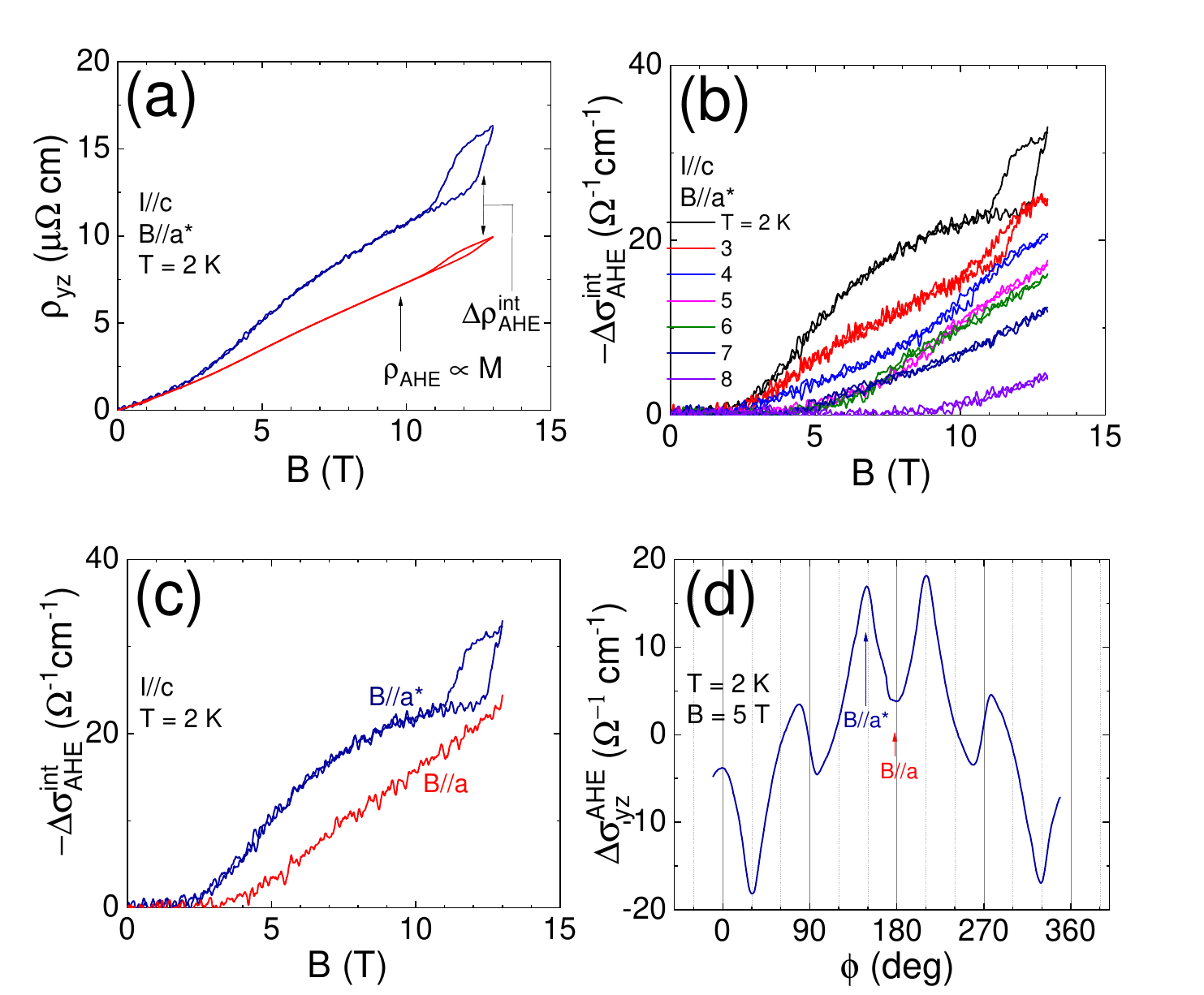}
	\caption{\label{figint} 
		{Intrinsic anomalous Hall conductivity and its angular dependence.}
		(a)  Linear and non-linear  contribution of anomalous Hall resistivity with $B \parallel a^*$. The blue (red) curves indicate the total ( linear in M) anomalous Hall resistivity. The difference between blue and red curves represents the  non-linear intrinsic contribution. (b) The temperature dependence of the intrinsic anomalous Hall conductivity after subtracting the extrinsic contribution with  $B \parallel a^*$. (c) The intrinsic anomalous Hall conductvity with $B \parallel a$ and $B \parallel a^*$ at $T$ = 2 K. (d) In-plane angular dependence of  non-linear  anomalous Hall conductivity at $B$ = 5T. $\phi$=60n for $B \parallel a$ and $\phi$=60n + 30 for $B \parallel a^*$. The current was applied parallel to $c$-axis.
	}
\end{figure}


After subtracting the linear-in-$M$ component, the remaining non-linear intrinsic AHC at different temperatures is shown in Fig. \ref{figint} (b). The AHC with different orientations is shown in (c). When $B \parallel a$ the intrinsic non-linear anomalous Hall conductivity becomes finite above $B_a$ $\simeq$ 4-5 T. When $B \parallel a^*$ the non-linear AHC emerges above $B_{a^*}$ $\simeq$ 2-3 T, and the AHC  $\Delta\sigma_{yz}^{int}=\frac{-\Delta\rho_{yz}^{int}}{\rho_{yy}\rho_{zz}+\rho_{yz}^2}$ reaches about 34 $\Omega^{-1} $cm$^{-1}$ at 13 T, comparable to ferromagnetic materials. 

Given the non-linear Berry curvature with field magnitude, we also explore the angle dependence of the AHE. The fact that the onset field for the non-linear AHC is almost double for $B\parallel a$ than for $B \parallel a^*$ indicates the Berry curvature can be sensitively tuned by rotating the magnetic field. Indeed, the intrinsic non-linear AHC as a function of tilt angle $\phi$, shown in Fig. \ref{figint} (d), is remarkably sensitive to the sample orientation. An overall two-fold oscillation is expected due to the reorientation of the magnetic structure with the applied magnetic field. However, at $B =$ 5 T, the absolute value of  AHC reaches local maxima when $B \parallel a^*$ (namely, 30, 90, 150, 210, 270 and 330 degrees). On the other hand, the absolute value of $\Delta\sigma^{int}_{AHE}$ almost vanishes at every 60 degrees, when $B \parallel a$. This demonstrates that the Berry curvature of our system can be highly tuned by small sample rotations. 


\begin{figure}[h]
	\includegraphics[width=\linewidth]{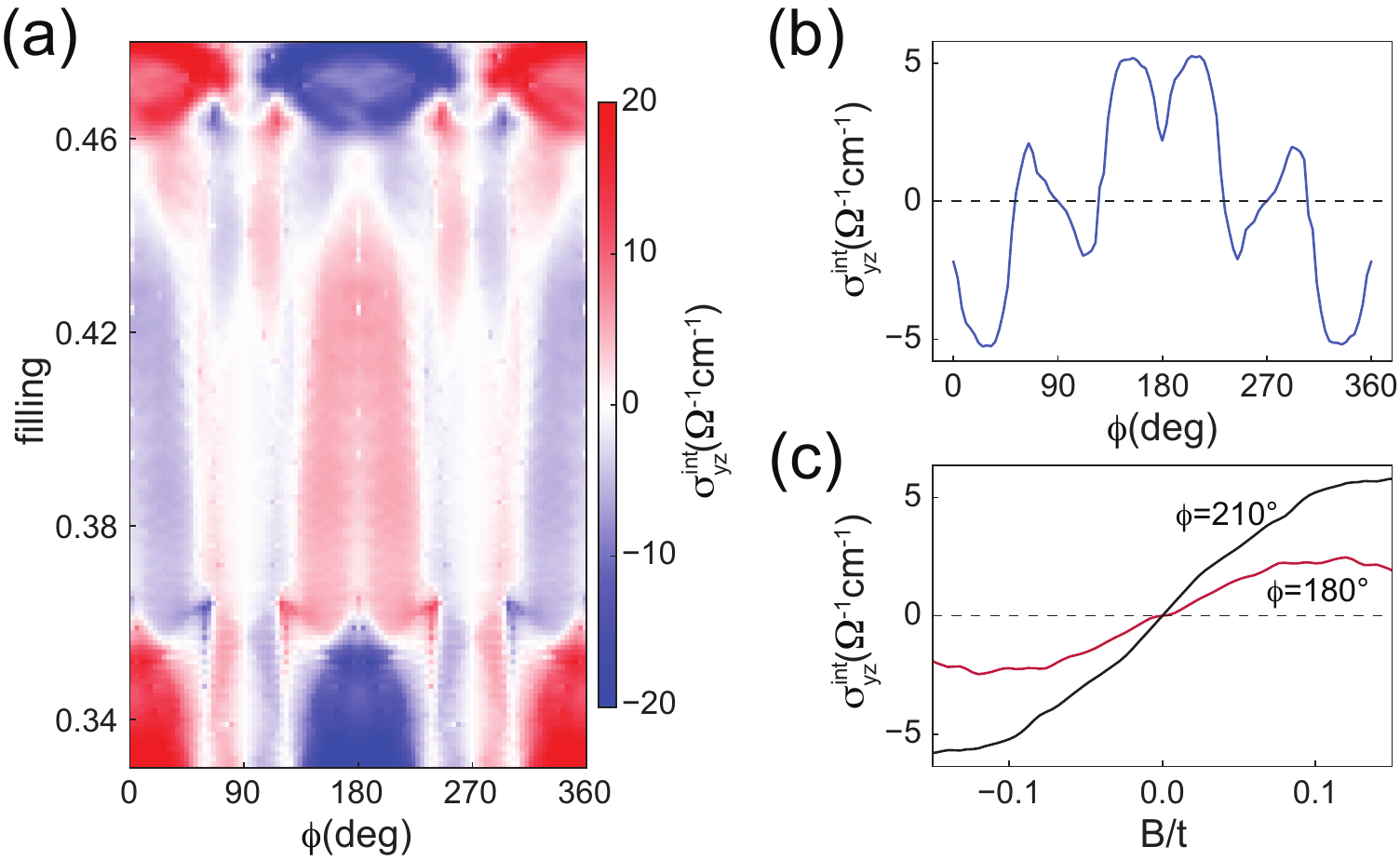}
	\caption{\label{figTheory} 
		{Theoretically calculated intrinsic anomalous Hall conductance in model Eq.~\ref{eqml} (color online). (a) The anomalous Hall conductance $\sigma_{yz}^{\rm int}$ as a function of the field angle $\phi$ and electron filling. (b) $\sigma_{yz}^{\rm int}$ vs  $\phi$ for fixed filling at 0.38.  (c) $\sigma_{yz}^{\rm int}$ as function of the field strength $B$ for the two field angles $\phi=180^\circ$ and $210^\circ$. Here we use the lattice parameter $a_1=0.862$ nm to obtain $\sigma_{yz}^{\rm int}$.}
	}
\end{figure}

{\bf Theoretical model}

To better understand the AHC in U$_3$Ru$_4$Al$_{12}$ with field and angle we construct a minimal model that captures the essential physics (see SI). The coplanar spin structure suggests that the intrinsic contribution to the AHE is a consequence of the Berry curvature created by the electronic structure in momentum space, as opposed to a real space contribution\cite{lee2009unusual,hirschberger2018skyrmion}. We introduce a Kondo lattice model to describe our system (see SI)
\begin{align}\label{eqml}
H = \sum_{\langle i\alpha,j\beta \rangle} t_{i\alpha,j\beta} c_{i\alpha}^\dagger c_{j\beta} - J\sum_{i\alpha} c_{i\alpha}^\dagger \bm{S}_{i\alpha} \cdot \bm{\sigma}  c_{i\alpha}\nonumber
\\- \sum_{i\alpha} c_{i\alpha}^\dagger \bm{B} \cdot \bm{\sigma}  c_{i\alpha}+ it_{i\alpha,j\beta}^{so}\sum_{\langle i\alpha, j\beta\rangle} c_{i\alpha}^\dagger \bm{n}_{\alpha\beta} \cdot \bm{\sigma} c_{j\beta},
\end{align}
where $c_{i\alpha}=(c_{i\alpha\uparrow},c_{i\alpha\downarrow})^\top$ is the conduction electron annihilation operator of a two-component spinor at the $i$-th unit cell and of the $\alpha$ sublattice. The localized $f$ electrons are responsible for the magnetic moments $\bm{S}_i$, which are treated classically. We take the experimentally measured spin configuration $\bm{S}_i$. \cite{troc2012single} The spin-orbit coupling (SOC) vectors $\bm{n}_{\alpha\beta}$ respect the $\mathcal{T}$ and $\mathcal{P}$ symmetries, and are defined in SI \cite{SI}. As discussed in detail in the Supplement \cite{SI}, we need to break $\mathcal{PT}$, $\{\mathcal{P}\mathcal{C}_{2z}|(0,0,1/2)\}$ and $\{\mathcal{T}\mathcal{M}_{z}|(0,0,1/2)\}$ symmetries to have a nonzero AHC. Here $\mathcal{C}_{2z}$  is the two-fold rotation of the lattice with respect to the rotation axis along the $z$ direction, $\mathcal{M}_z$ denotes the mirror symmetry with the mirror plane perpendicular to the $z$ direction, and $(0,0,1/2)$ denotes a nonprimitive translation along the $z$ direction by half a lattice constant. The $\bm{B}$ field breaks $\mathcal{PT}$ and the SOC breaks $\{\mathcal{P}\mathcal{C}_{2z}|(0,0,1/2)\}$ and $\{\mathcal{T}\mathcal{M}_{z}|(0,0,1/2)\}$ symmetries.

We calculate the intrinsic anomalous Hall conductivity $\sigma_{yz}^{\rm int}$ as a function of field angle $\phi$ and electron filling, see Fig.~\ref{figTheory} (a). In a range of filling, from about 0.36 to 0.42, the theoretically calculated $\sigma_{yz}^{\rm int}$ is consistent with the experimental results. The field strength and direction dependence of $\sigma_{yz}$ at one typical filling at 0.38 are displayed in Figs.~\ref{figTheory} (b) and (c). The simple model captures semi-quantitatively the experimental observation in Figs.~\ref{figint} (d) and (c).

\begin{figure}[h]
	\includegraphics[width=\linewidth]{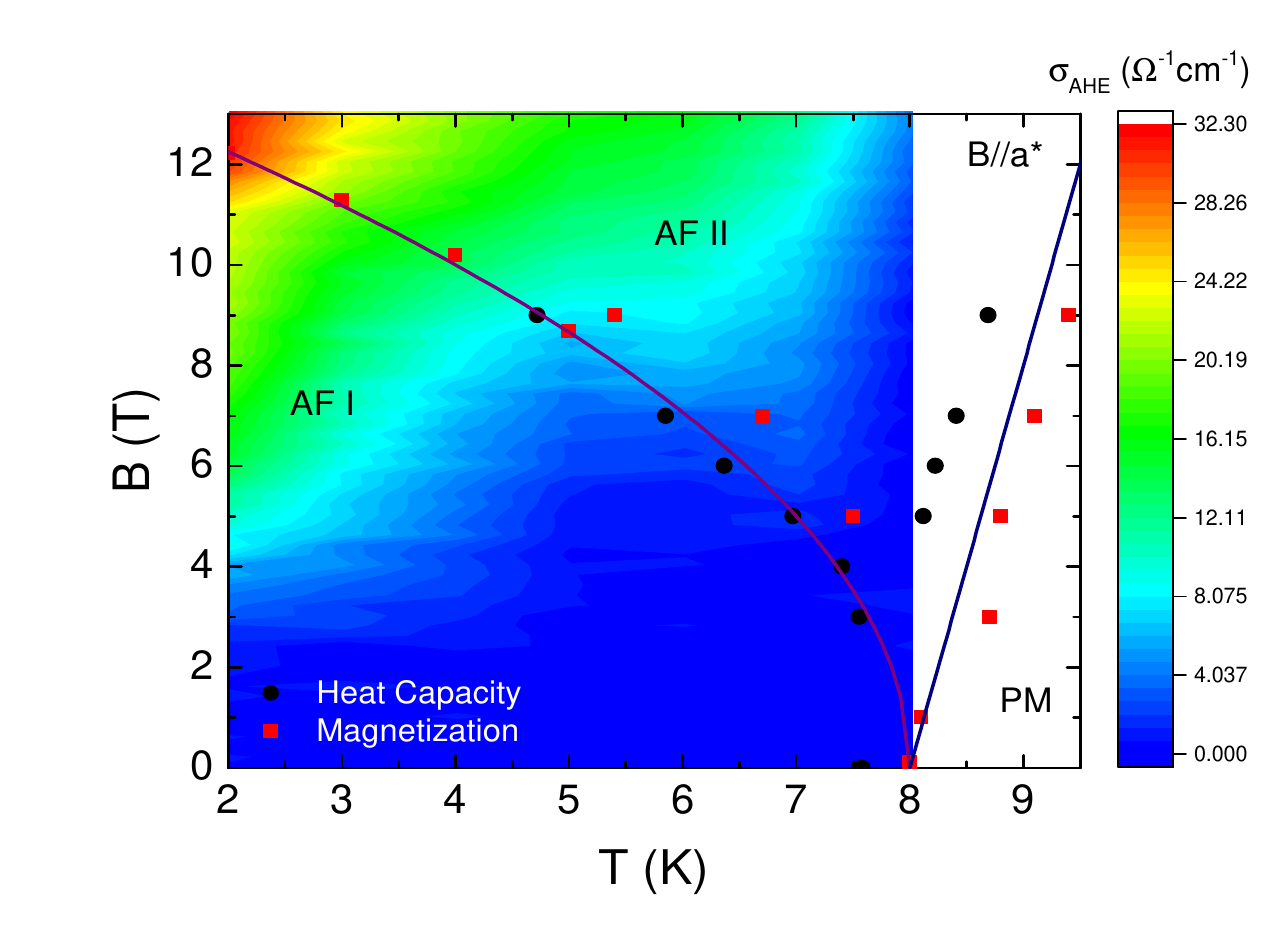}
	\caption{\label{figPhase} 
		{Magnetic and topological phase diagram of U$_3$Ru$_4$Al$_{12}$ (color online).}
		The magnetic field is applied parallel to $a^*$. The magnetic phase transitions are determined by heat capacity and magnetization measurements, which are consistent with MR. The contour plot indicates the amplitude of the  non-linear AHC.
	}
\end{figure}

{\bf Discussion}

In Fig. \ref{figPhase}, we show the magnetic phase diagram and non-linear AHC contour plot of U$_3$Ru$_4$Al$_{12}$. With the magnetic field $B$  applied parallel to $a^*$, a magnetic phase transition is observed in  MR, magnetization and heat capacity measurements (see SI), which is consistent with a recent study\cite{gorbunov2019magnetic}. At low fields, the system stays in the antiferromagnetic phase I (AF I). At high fields, there is a metamagnetic transition and the system enters antiferromagnetic phase II (AF II). Surprisingly, the AHC behaves almost independently from the magnetic phases. At low temperatures, a magnetic field $B_{a^*}$ required to push the Berry curvature to a non-linear regime is almost constant, and much smaller than $B_M$. At high temperatures, $B_{a^*}$ increases rapidly as the temperature approaches $T_N$, while $B_M$ decreases. Neutron scattering measurements at high fields would be helpful to understand the magnetic structure of the AF II phase and the evolution of the AHC.

In recent reports, the sign of the Berry curvature was switched by flipping the domain structure\cite{surgers2014large,ye2018massive,liu2018giant,nayak2016large,nakatsuji2015large}, but it is quite rare to observe an in-plane Berry-curvature switching without inducing a magnetic transition. This is reasonable because in most cases the in-plane magnetic anisotropy is quite small, as is also the case in U$_3$Ru$_4$Al$_{12}$. What causes such an anisotropic and field-dependent behavior of Berry curvature in U$_3$Ru$_4$Al$_{12}$? A non-coplanar spin texture could generate such an effect, but would require the coplanar spin arrangement in zero field to cant out of the plane. Our theoretical model demonstrates that a field and angular dependent momentum-space Berry curvature could arise in U$_3$Ru$_4$Al$_{12}$ if the strength of the field becomes a sizable fraction of the bandwidth.  For most materials this is not possible with today's magnets, but the renormalized bandwidth found in heavy fermion materials enables this mechanism as a result of strong electronic correlations.

{\bf Conclusions} 

We have demonstrated that a heavy fermion non-collinear antiferromagnet can be driven into a regime with a non-linear response of the Berry curvature.  A similar field-induced Berry curvature was also observed in the non-collinear antiferromagnet and attributed to the proximity of Weyl nodes to the Fermi energy \cite{suzuki2016large}. The total AHC in U$_3$Ru$_4$Al$_{12}$ reaches 0.21 e$^2$/$ha$ for $\sigma_{yz}$ where $a$ is the a-axis lattice parameter of which a minimum of 0.08 e$^2$/$ha$ can be attributed to an intrinsic Berry curvature effect. For $\sigma_{xy}$, an even larger value of total AHC (0.68 e$^2$/$hc$) was reached. ($c$ is the c-axis lattice parameter). These large values are found  despite U$_3$Ru$_4$Al$_{12}$ being a 3-dimensional electron system.  The combination of large effective masses, crystal electric fields, and a frustrated distorted kagom$\acute{e}$ lattice enables this system to possess strong sensitivity of the Berry curvature to the magnitude and direction of an applied field and sample rotation, further illustrating that $f$-electron systems are fruitful playgrounds to explore tuning of their Berry curvature.

%
%
%
%

{\bf Acknowledgement} 
We thank V. Zapf for assistance with measurements at the National High Magnetic Field Laboratory. The experimental work was supported by the U.S. Department of Energy, Office of Science, Basic Energy Sciences, Materials Sciences and Engineering Division. The theoretical work was carried out under the auspices of the U.S. DOE Award No. DE-AC52-06NA25396 through the LDRD program, and was supported by the Center for Nonlinear Studies at LANL. A portion of this work was performed at the National High Magnetic Field Laboratory user facility, which is supported by the National Science Foundation Cooperative Agreement No. DMR-164479 the state of Florida and the U.S. Department of Energy.


%

\bibliography{RefU3Ru4Al12}

\end{document}



\title{Large Tunable Anomalous Hall Effect in Kagom$\acute{e}$ Antiferromagnet U$_3$Ru$_4$Al$_{12}$}
\author{T. Asaba$^{1*}$, Ying Su$^{2}$, M. Janoschek$^{1,3}$, J. D. Thompson$^{1}$, S. M. Thomas$^{1}$,  E. D. Bauer$^1$, Shi-Zeng Lin$^{2}$, F. Ronning$^{4*}$}

\affiliation{
	$^1$MPA-CMMS, Los Alamos National Laboratory, NM, 87545 USA \\
	$^2$Theoretical Division, T-4 and CNLS, Los Alamos National Laboratory, NM, 87545, USA \\
	$^3$Laboratory for Scientific Developments and Novel Materials, Paul Scherrer Institute, Switzerland \\
	$^4$Institute for Materials Science, Los Alamos National Laboratory, NM, 87545, USA}
\date{\today}
\begin{abstract}

\end{abstract}

\maketitle
\section{Transport measurement}
\begin{figure}[h]
	\includegraphics[width=\linewidth]{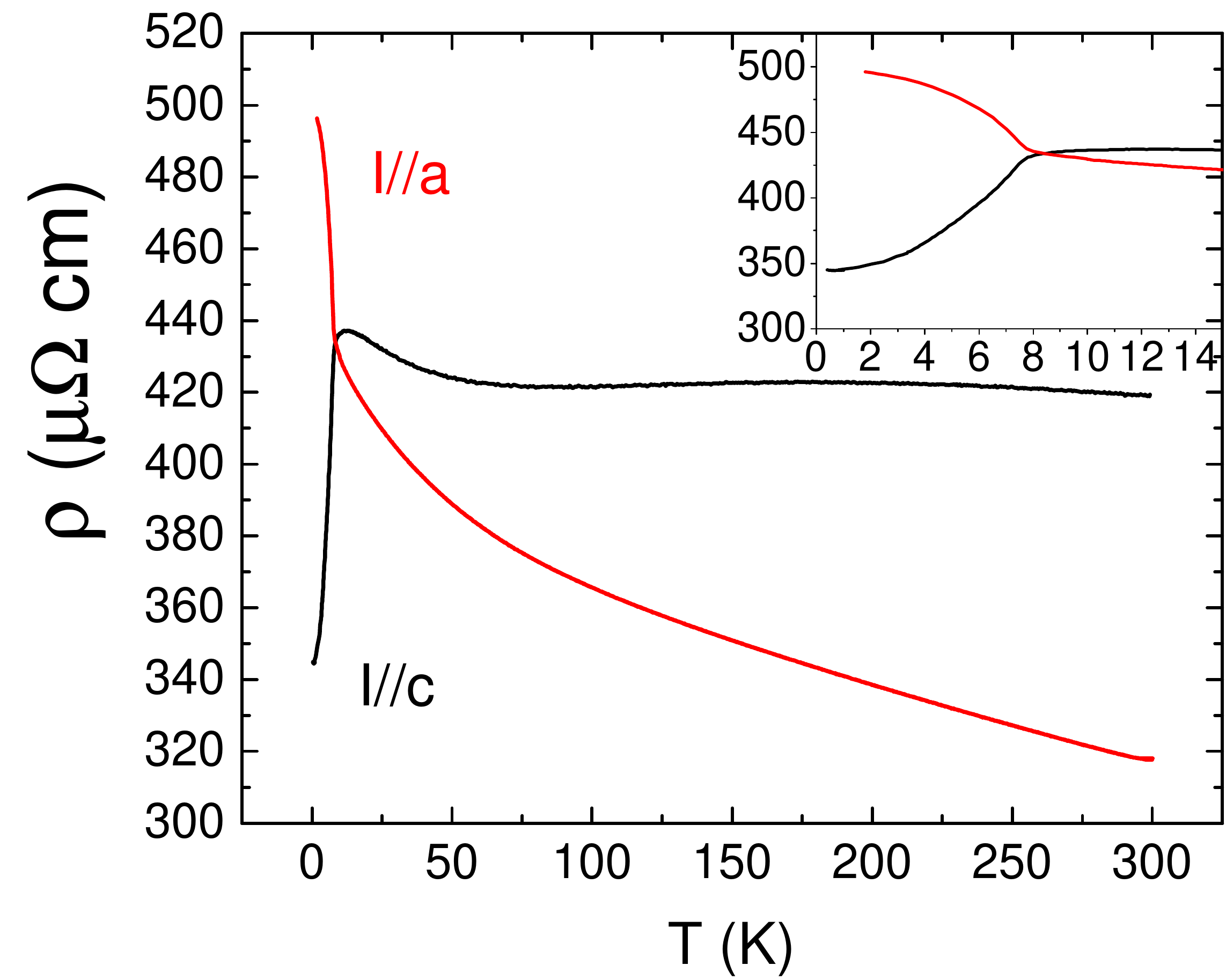}
	\caption{\label{RT} 
		{(color online).} Temperature dependence of resistivity for $I \parallel a$ and $I \parallel c$. Inset is the expanded plot around $T = T_N$.
	}
\end{figure}

Fig. \ref{RT}  shows the temperature dependence of $\rho_{xx}$ and $\rho_{zz}$ at zero fields. The poor metal behavior in both $I\parallel a$ and $I\parallel c$ configurations is consistent with the previous study\cite{troc2012single}. The resistivity shows a significant drop or enhancement at the N$\acute{e}$el temperature $T_N$ = 8 K (inset).

\begin{figure}[h]
	\includegraphics[width=\linewidth]{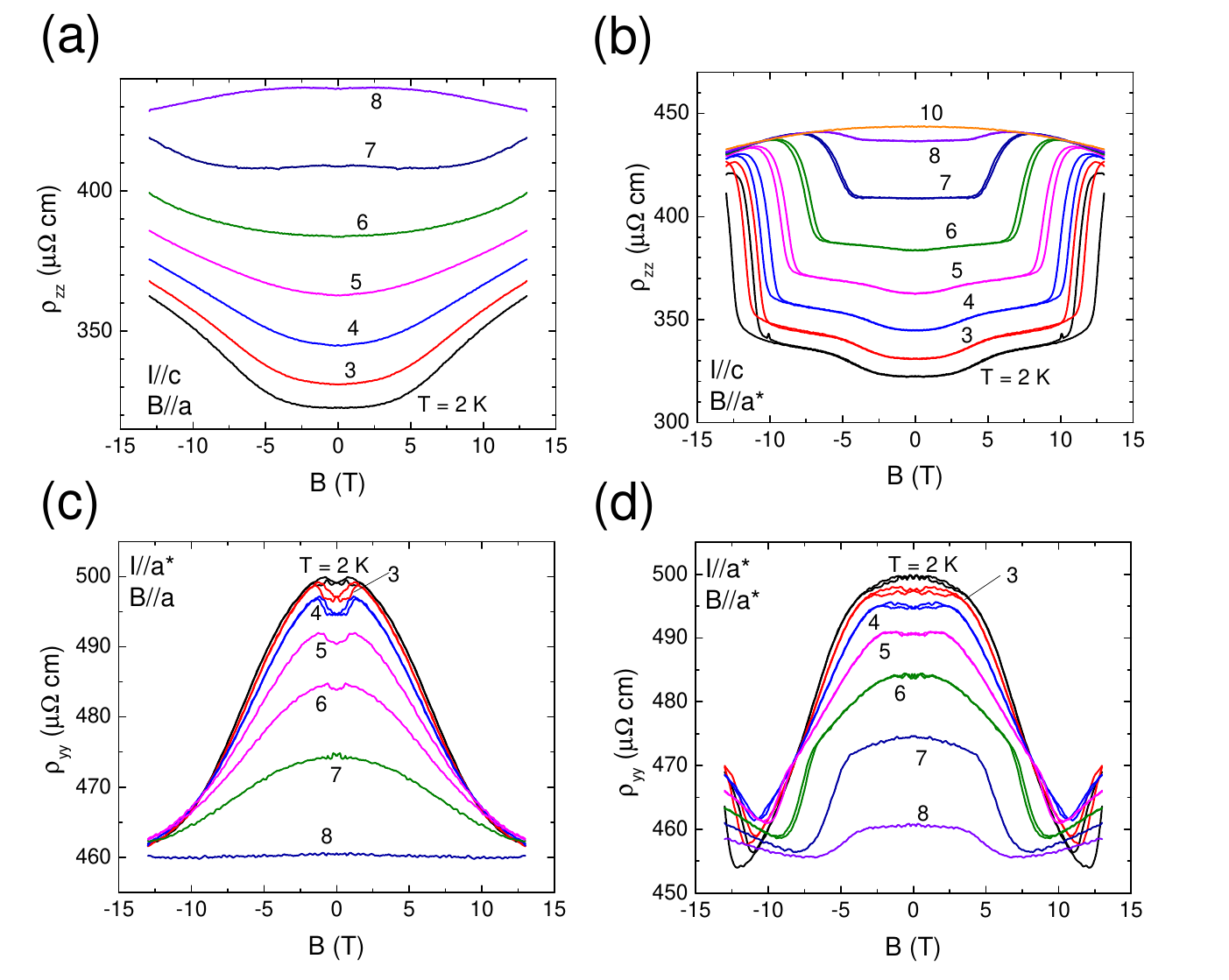}
	\caption{\label{MR} 
		{(color online).} (a)(b)Temperature dependence of magnetoresistance $\rho_{zz}$ for (a) $B \parallel a$ and (b) $B \parallel a^*$. (c)(d) Temperature dependence of magnetoresistance $\rho_{yy}$ for (c) $B \parallel a$ and (d) $B \parallel a^*$.
	}
\end{figure}

\begin{figure}[h]
	\includegraphics[width=\linewidth]{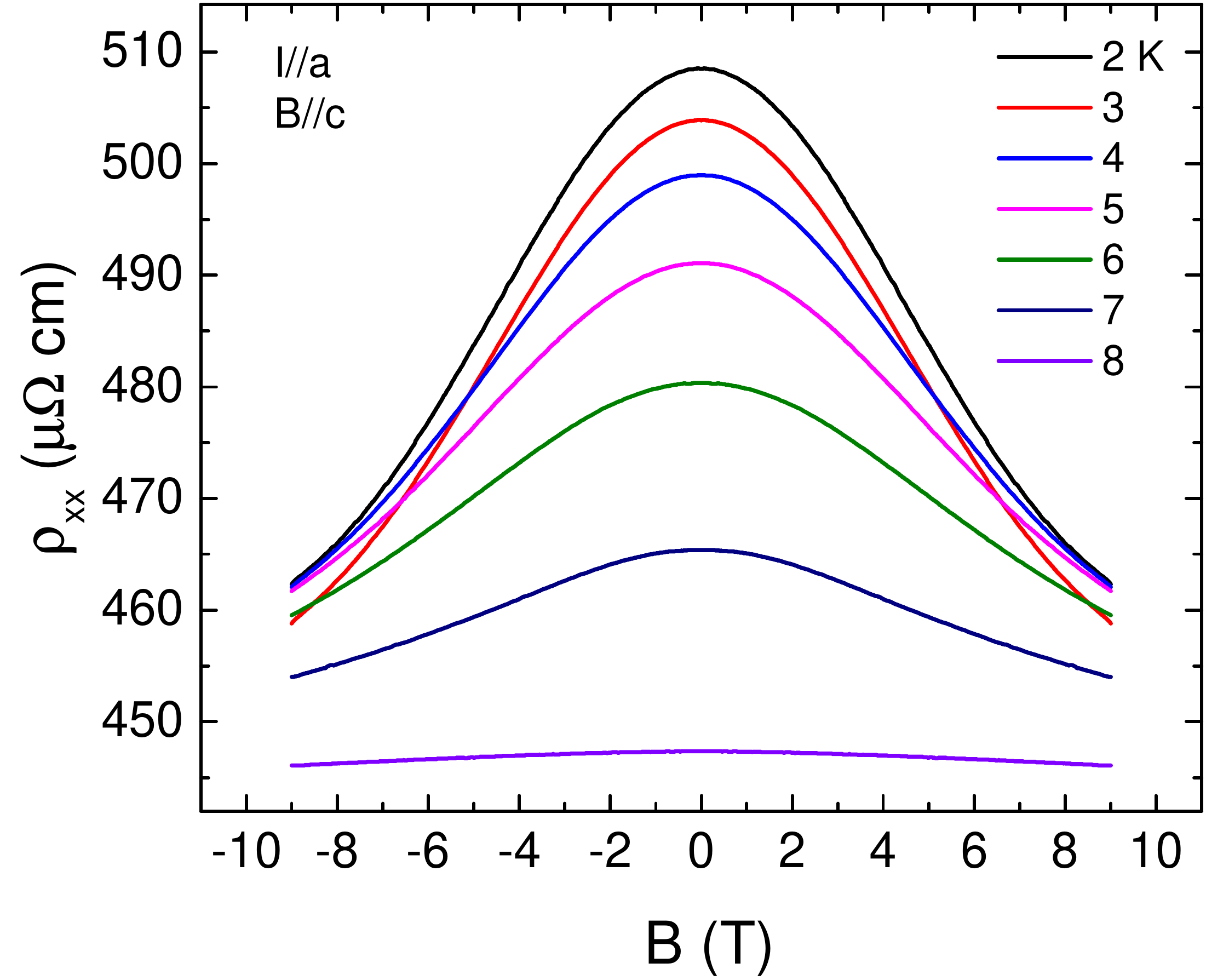}
	\caption{\label{MRc} 
		{(color online). Magnetoresistance with I//a and B//c.} 
	}
\end{figure}

Fig. \ref{MR} (a)(b) displays the temperature dependence of in-plane magnetoresistance (MR) for the U$_3$Ru$_4$Al$_{12}$ single crystal with $I\parallel c$. Interestingly, while the crystal has a layered hexagonal structure, the in-plane anisotropy is substantial. In the $B\parallel a$ configuration (shown in (a)), at low temperatures a flat MR was observed up to a few tesla, after which the MR becomes positive. On the other hand, with $B\parallel a^*$ configuration (shown in (b)), at 2 K a clear hysteresis loop with large MR enhancement reveals a first order phase transition  around the critical magnetic field $B_M$ = 12 T. As the temperature increases, $B_M$ decreases and completely vanishes above $T_N$. The phase transition is also observed in $\rho_{yy}$ as shown in (d), although the overall MR is negative instead of positive. No sign of a metamagnetic transition was observed with B$\parallel$c, either, as shown in Fig. \ref{MRc}.

\begin{figure}[h]
	\includegraphics[width=\linewidth]{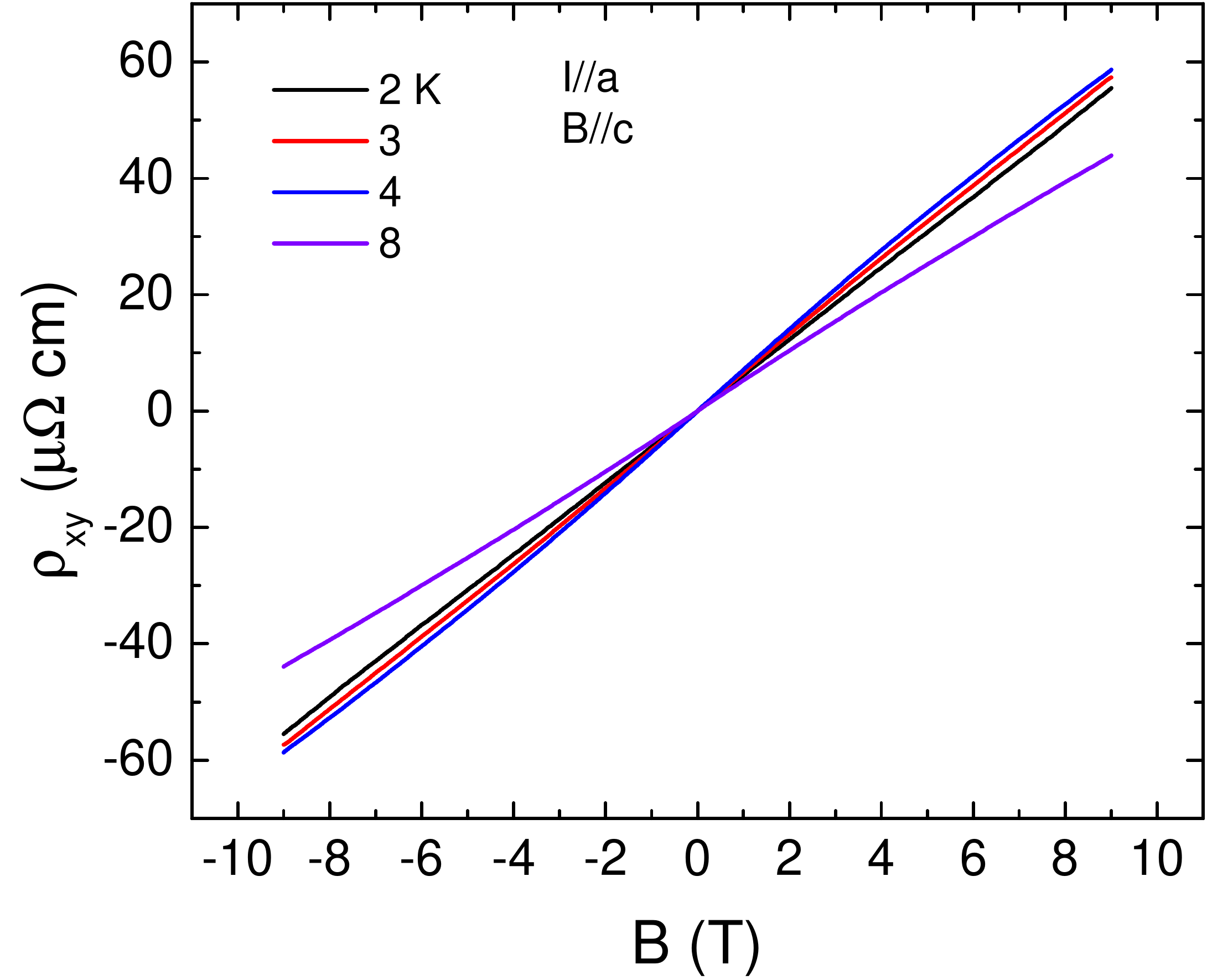}
	\caption{\label{Hallc} 
		{(color online).The Hall resistivity $\rho_{xy}$ at different temperatures. The magnetic field is applied parallel to the $c$-axis.} 
	}
\end{figure}

Since the magnetic structure is co-planar in the basal plane, one would expect that an out-of-plane magnetic field breaks the mirror symmetry, resulting in a large anomalous Hall conductivity (AHC). Indeed, this is the case as shown in Fig. \ref{Hallc}. The out-of-plane AHC reaches almost double the in-plane one.

\begin{figure}[h]
	\includegraphics[width=\linewidth]{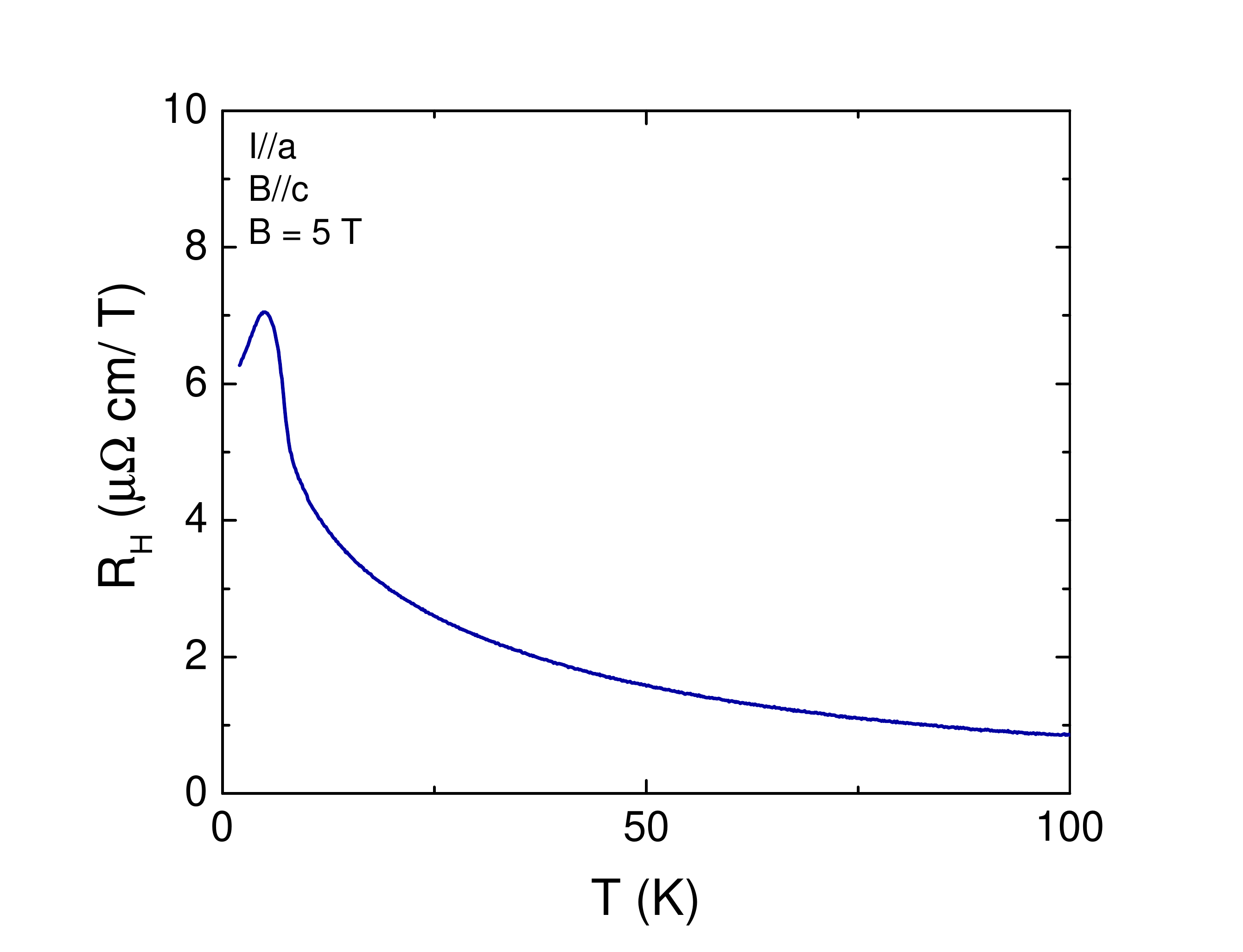}
	\caption{\label{RxyT} 
		{(color online). The Hall coefficient R$_H$ = $\rho_{xy}$/$B$ as a function of temperature. The applied field is $B$ = 5 T. } 
	}
\end{figure}

The contribution from the ordinary Hall could be estimated by plotting the Hall coefficient R$_H$ = $\rho_{xy}$/B as a function of temperature as shown in Fig. \ref{RxyT}. At high temperatures, the AHC shrinks, from which one may infer that the ordinary Hall effect is vanishingly small. This is consistent with the large carrier density inferred from the large Sommerfelt coefficient $\gamma$ shown below.  If the system is in fact a compensated metal a large (and potentially non-linear in magnetic field) contribution to the ordinary Hall effect could arise at low temperatures if the mobility were highly temperature dependent. The essentially temperature independent resistivity, however, rules out this possibility.

\section{Heat capacity measurement}

\begin{figure}[h]
	\includegraphics[width=\linewidth]{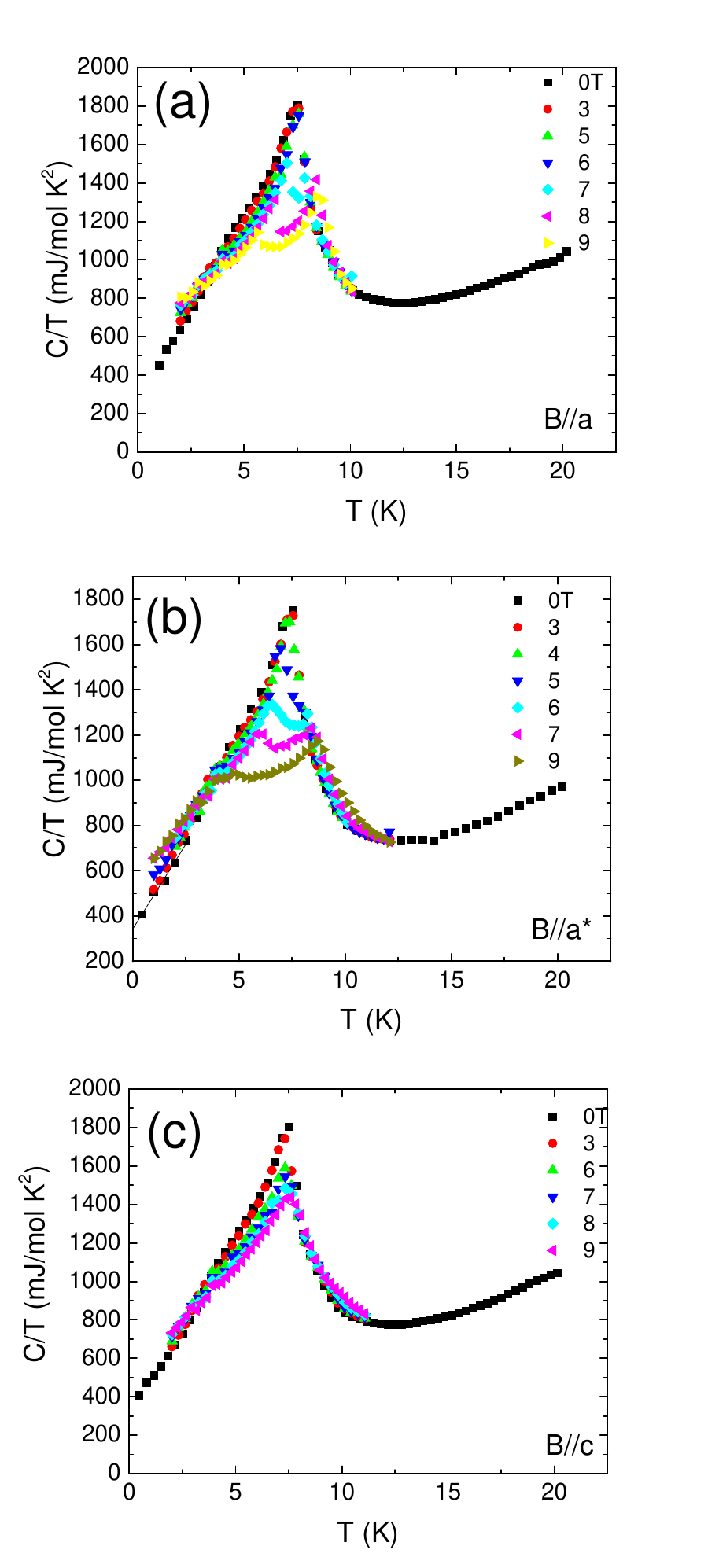}
	\caption{\label{FigHC} 
		{(color online). Heat capacity data with different magnetic field orientations. } 
		}
\end{figure}

Heat capacity measurements with different field orientation are shown in Fig. \ref{FigHC} (a)-(c).  The estimated value of Sommerfelt coefficient $\gamma$ is ~110 mJ/mol-U K$^2$, indicating a moderately heavy fermion behavior. With $B\parallel$a*, there are two phase transitions observed, consistent with the transport results. Interestingly, similar but narrower phase transitions were observed with $B\parallel$a, but restricted to higher temperatures. No phase transition except for the N$\acute{e}$el transition was observed with $B\parallel$c, which is also consistent with the transport data.

\section{Magnetization measurement}

\begin{figure}[h]
	\includegraphics[width=\linewidth]{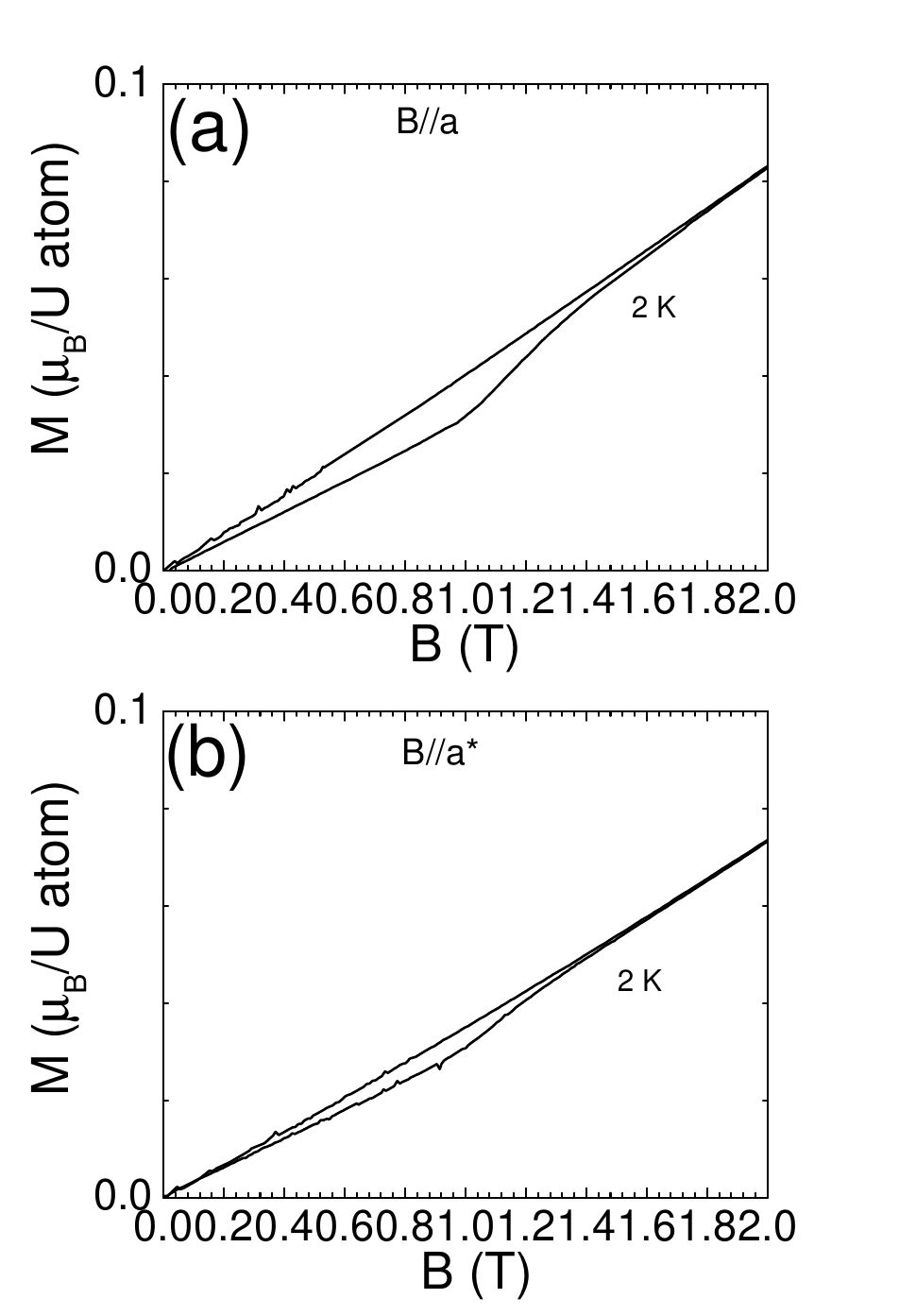}
	\caption{\label{FigMHLoop} 
		{(color online). Expanded magnetization data as a function of magnetic field showing hysteresis loops for (a) $B \parallel a$ and (b) $B \parallel a^*$. } 
	}
\end{figure}

The expanded data at $T$ = 2K is shown in Fig. \ref{FigMHLoop}. Hysteresis loops are observed about zero field in both $B \parallel a$ and $B \parallel a^*$ orientations suggesting the reorientation of magnetic domains in low magnetic fields..

\section{Theoretical model}

\begin{figure}[b]
	\includegraphics[width=\linewidth]{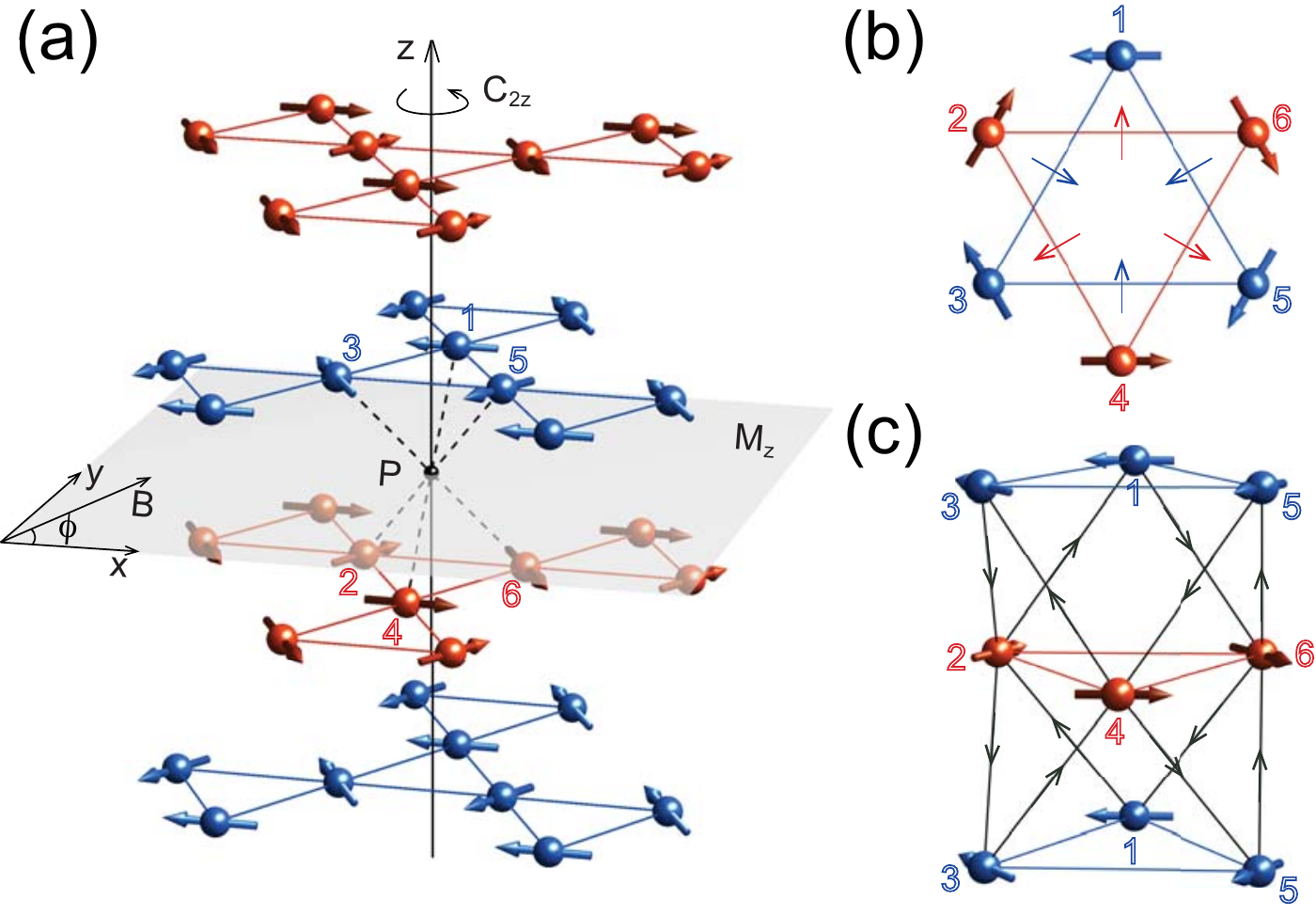}
	\caption{\label{figs1} 
		{(color online).} 
		(a) The 3D kagom$\acute{e}$ lattice formed by the uranium atoms of U$_3$Ru$_4$Al$_{12}$. The non-collinear magnetic texture is indicated by the arrows on the uranium atoms. The inversion center, mirror plane, and two-fold rotation axis are respectively marked by $P$, $M_z$, and $C_{2z}$. The six atoms belonging to different sublattices in a unit cell are labeled from 1 to 6. (b) The coplanar arrows perpendicular to the NN intralayer bonds denote the unit vectors $\bm{n}_{13}$, $\bm{n}_{24}$, $\bm{n}_{35}$, $\bm{n}_{46}$, $\bm{n}_{51}$, and $\bm{n}_{62}$ in the NN intralayer SOC. (c) The black bonds denote the NN interlayer hopping. The arrows along the interlayer bonds stand for the unit vectors $\bm{n}_{12}$, $\bm{n}_{32}$, $\bm{n}_{34}$, $\bm{n}_{54}$, $\bm{n}_{16}$,  and $\bm{n}_{56}$ in the NN interlayer SOC.  	}
\end{figure}

To understand the dependence of the magnetic field on the anomalous Hall effect (AHE) in U$_3$Ru$_4$Al$_{12}$ we construct a minimal model to capture the essential physics. Here we consider only the magnetic uranium atoms that form kagom$\acute{e}$ lattices stacking along the $z$ direction, as shown in Fig. \ref{figs1}(a). Local spins on the uranium atoms constitute a  non-collinear  magnetic structure according to the neutron scattering study \cite{PhysRevB.85.064412} (see Fig.~\ref{figs1}).  Conduction  electrons hopping on the 3D kagom$\acute{e}$ lattice are coupled to the  magnetic structure through the local exchange interaction and is described by the Kondo lattice Hamiltonian 
\begin{equation}
H_0 = \sum_{\langle i\alpha,j\beta \rangle} t_{i\alpha,j\beta} c_{i\alpha}^\dagger c_{j\beta} - J\sum_{i\alpha} c_{i\alpha}^\dagger \bm{S}_{i\alpha} \cdot \bm{\sigma}  c_{i\alpha},
\end{equation}
where the first term describes both the nearest-neighbor (NN) intralayer and interlayer hopping, and the second term is the local exchange interaction. Here $c_{i\alpha}=(c_{i\alpha\uparrow},c_{i\alpha\downarrow})^\top$ is the conduction electron  annihilation operator of a two-component spinor at the $i$-th unit cell and of the $\alpha$ sublattice. In the 3D kagom$\acute{e}$ lattice, each unit cell contains 6 atoms labeled as $\alpha=1$ to  $6$, as shown in Fig. \ref{figs1}. $t_{i\alpha, j\beta}=t$  for the NN intralayer hopping, while $t_{i\alpha,j\beta}=t_\perp$ for the NN interlayer hopping. The localized $f$ electrons are modeled by magnetic moments $\bm{S}_{i\alpha}$ with $|\bm{S}_{i\alpha}|=1$, which are treated classically.  $\bm{S}_{i\alpha}$ encodes the local non-collinear  magnetic texture, and $\bm{\sigma}$ is the vector of Pauli matrices. Here we neglect the distortion of kagom$\acute{e}$ lattice and deem all the bonds of kagom$\acute{e}$ lattice having the same length to simplify the model. This simplification does not affect the symmetry consideration of the system.

In order to elucidate in which circumstance the  AHE can occur, we analyze the symmetry of the model.  The 3D kagom$\acute{e}$ lattice has inversion symmetry and the inversion center is shown in Fig. \ref{figs1}(a). Because any two inversion partners have opposite spins, all the local spins must be reversed after the inversion operation.  Therefore, the system is invariant under the combined inversion and time-reversal operation since the latter flips all spins back. The $\mathcal{PT}$ symmetry of the system ensures the Berry curvature $\Omega_{\alpha\beta}^{(n)}(\bm{k})=-\Omega_{\alpha\beta}^{(n)}(\bm{k})=0$, where $\Omega_{\alpha\beta}^{(n)}(\bm{k})=\partial_{k_\alpha}A_{\beta}^{(n)}(\bm{k})-\partial_{k_\beta}A_{\alpha}^{(n)}(\bm{k})$ and $A_{\alpha}^{(n)}(\bm{k})=i\bra{\psi_n(\bm{k})}\partial_{k_\alpha}\ket{\psi_n(\bm{k})}$ is the Berry connection of the $n$th eigenstate $\ket{\psi_n(\bm{k})}$. Namely, the intrinsic AHC is zero since
\begin{equation}
\sigma_{\alpha\beta}^{\rm int} = -\frac{e^2}{\hbar} \sum_n \int \frac{d^3 k}{(2\pi)^3} f(E_n(\bm{k})-\mu) \Omega^{(n)}_{\alpha\beta}(\bm{k}),
\end{equation}
where $\mu$ is the chemical potential and $f(E_n(\bm{k})-\mu)$ is the Fermi function.

\begin{figure}[h]
	\includegraphics[width=\linewidth]{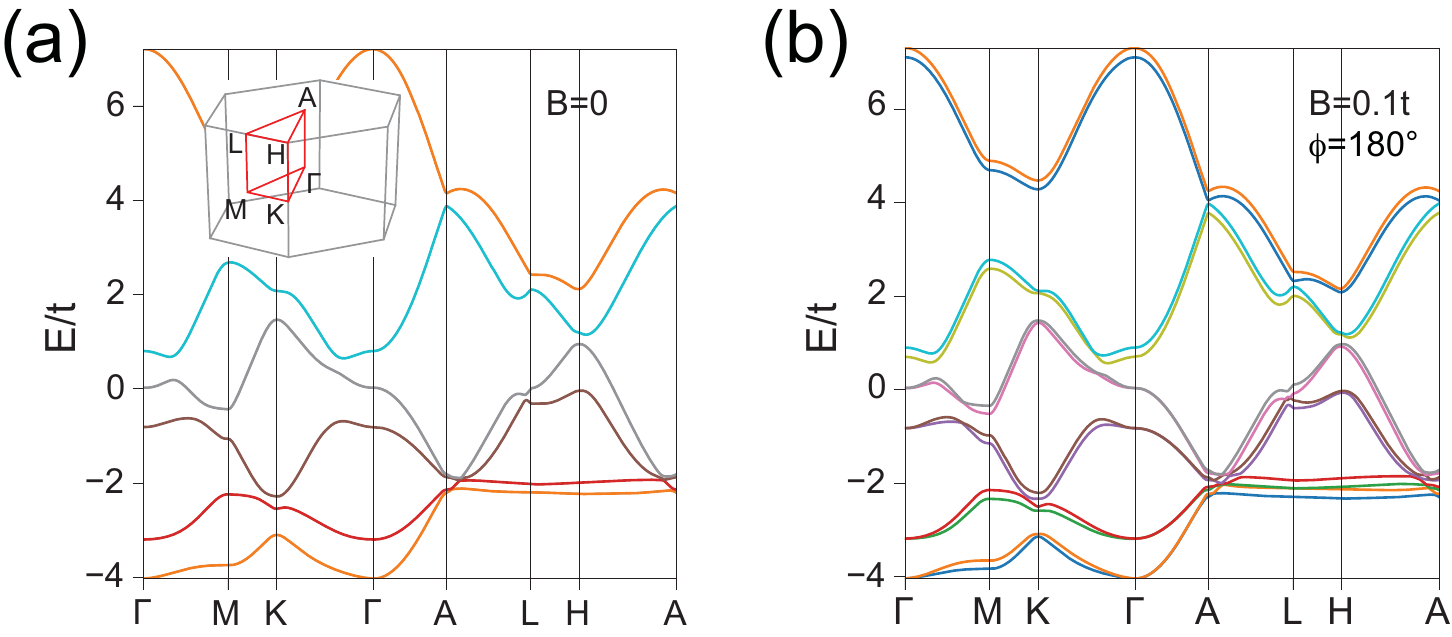}
	\caption{\label{figs2} 
		{(color online). The energy spectra along the high symmetry path in the inset of (a). (a) The spectrum in the absence of Zeeman field and all the energy bands are doubly degenerate. (b) The spectrum for $B=0.1t$ and $\phi=180^\circ$.} 
	}
\end{figure}

The Zeeman interaction can break the $\mathcal{PT}$ symmetry. In experiments, an in-plane magnetic field is applied and the Zeeman interaction is depicted by the Hamiltonian
\begin{equation}
H_z=- \sum_{i\alpha} c_{i\alpha}^\dagger \bm{B} \cdot \bm{\sigma}  c_{i\alpha},
\end{equation}
where $\bm{B}=(B\cos\phi,B\sin\phi,0)$ denotes the in-plane Zeeman field and the polar angle $\phi$ is defined in Fig. \ref{figs1}(a). Here we assumed that the field does not alter the spin texture. This assumption is consistent with the experimental observation $\sigma_{yz}^{\rm int} (\bm{B})=-\sigma_{yz}^{\rm int} (-\bm{B})$. This is further supported by the fact that no anomaly was observed in magnetization and specific heat data until the metamagnetic transition.

There are two extra nonsymmorphic symmetries that render the AHE	invisible. First, the system has the symmetry $\{\mathcal{P}\mathcal{C}_{2z}|(0,0,1/2)\}$ where $\mathcal{C}_{2z}$  is the two-fold rotation of the lattice with respect to the rotation axis along the $z$ direction in Fig. \ref{figs1}(a), and $(0,0,1/2)$ denotes a nonprimitive translation along the $z$ direction by half of a lattice constant. This symmetry results in
\begin{equation}
\begin{split}
& \Omega_{xy}^{(n)}(k_x,k_y,k_z) = \Omega_{xy}^{(n)}(k_x,k_y,-k_z), \\
& \Omega_{yz}^{(n)}(k_x,k_y,k_z) = -\Omega_{yz}^{(n)}(k_x,k_y,-k_z), \\
& \Omega_{zx}^{(n)}(k_x,k_y,k_z) = -\Omega_{zx}^{(n)}(k_x,k_y,-k_z). 
\end{split}
\end{equation}
Additionally, the system has the $\{\mathcal{TM}_z|(0,0,1/2)\}$ symmetry, where $\mathcal{M}_z$ denotes the mirror symmetry with the mirror plane perpendicular to the $z$ direction as shown in Fig. \ref{figs1}(a). Due to this symmetry, the Berry curvature follows
\begin{equation}
\begin{split}
& \Omega_{xy}^{(n)}(k_x,k_y,k_z) = -\Omega_{xy}^{(n)}(-k_x,-k_y,k_z), \\
& \Omega_{yz}^{(n)}(k_x,k_y,k_z) = \Omega_{yz}^{(n)}(-k_x,-k_y,k_z), \\
& \Omega_{zx}^{(n)}(k_x,k_y,k_z) = \Omega_{zx}^{(n)}(-k_x,-k_y,k_z). 
\end{split}
\end{equation}
The nonsymmorphic symmetries $\{\mathcal{P}\mathcal{C}_{2z}|(0,0,1/2)\}$ and $\{\mathcal{TM}_z|(0,0,1/2)\}$ together also ensure the Berry curvature $\Omega_{\alpha\beta}^{(n)}(\bm{k})=-\Omega_{\alpha\beta}^{(n)}(-\bm{k})$ is an odd function in momentum space and hence no intrinsic AHE is expected.

The AHE can be realized when $\{\mathcal{P}\mathcal{C}_{2z}|(0,0,1/2)\}$ or $\{\mathcal{TM}_z|(0,0,1/2)\}$ symmetry is broken. Especially, in the experiments, $\sigma_{yz}^{\rm int} \neq0$ indicates $\{\mathcal{P}\mathcal{C}_{2z}|(0,0,1/2)\}$ must be violated. To break these symmetries, we further consider the spin-orbit coupling (SOC) in our model. According to the experiments, the zero AHC in the absence of magnetic field indicates the SOC should respect the $\mathcal{PT}$ symmetry. Under this constraint, we introduce the SOC \cite{PhysRevLett.112.017205}
\begin{equation}
H_{so} = it_{i\alpha,j\beta}^{so}\sum_{\langle i\alpha, j\beta\rangle} c_{i\alpha}^\dagger \bm{n}_{\alpha\beta} \cdot \bm{\sigma} c_{j\beta},
\end{equation}
where $t_{i\alpha,j\beta}^{so}=t^{so}$ for NN intralayer coupling and  $t_{i\alpha,j\beta}^{so}=t^{so}_\perp$ for NN interlayer coupling. $\bm{n}_{\alpha\beta}=-\bm{n}_{\beta\alpha}$ are a set of unit vectors joining two NN intralayer or interlayer lattice sites as shown in Figs. \ref{figs1}(b) and \ref{figs1}(c).  In this case, the SOC breaks  both the $\{\mathcal{P}\mathcal{C}_{2z}|(0,0,1/2)\}$ and $\{\mathcal{T}\mathcal{M}_{z}|(0,0,1/2)\}$ symmetry, while respecting the $\mathcal{PT}$ symmetry as expected.

We consider all these interactions together as $H=H_0+H_z+H_{soc}$. To be concrete, we take $t_\perp=0.8t$, $t^{so}=0.3t$, $t^{so}_\perp=0.3t_\perp$, and $J=0.2t$. The energy spectra along the high symmetry path in the inset of Fig. \ref{figs2}(a)  are shown in Figs. \ref{figs2}(a) and \ref{figs2}(b) for $B=0$ and $B=0.1t$ with $\phi=180^\circ$, respectively. Apparently, all the energy bands are doubly degenerate in the absence of a Zeeman field due to $\mathcal{PT}$ symmetry, as shown in Fig. \ref{figs2}(a). The $\mathcal{PT}$ symmetry can be broken by the Zeeman interaction and each doubly degenerate bands in Fig. \ref{figs2}(a) splits into two as shown in Fig. \ref{figs2}(b).  

We calculate $\sigma_{yz}^{\rm int} $ as a function of field and angle over a wide range of electron filling. The field strength and direction dependence of $\sigma_{yz}^{\rm int} $ at one typical filling are displayed in Fig.~3 (b) and (c) in the main text. The simple model captures semi-quantitatively the experimental observation shown in Fig.~2 in the main text.

\bibliography{Refsm}